\documentclass[]{spie}  

\newcommand{\um}{$\mu$m}

\newcommand{\hh}{$\rm H_2$}
\newcommand{\hii}{H\,{\sc II}}
 
\usepackage{amsmath,amsfonts,amssymb}
\usepackage{graphicx}
\usepackage[table, x11names]{xcolor}
\usepackage[colorlinks=true, allcolors=blue]{hyperref}
\usepackage[rightcaption]{sidecap}

\title{Update on the slicer IFU for the Magellan InfraRed Multi-Object Spectrograph (MIRMOS)}

\author[a]{Maren Cosens}
\author[b]{Patricio Schurter}
\author[a]{Nicholas P. Konidaris II}
\author[a]{Gwen C. Rudie}
\author[a]{Andrew B. Newman}
\author[c]{Leon Aslan}
\author[d]{Robert Barkhouser}
\author[a]{Christoph Birk}
\author[e]{Julia Brady}
\author[a]{Tyson Hare}
\author[f]{Stephen C. Hope}
\author[a]{Charlie Hull}
\author[a]{Karim Kaismoune}
\author[a]{Daniel D. Kelson}
\author[a]{Gerrad Killion}
\author[g]{Alicia Lanz}
\author[h]{Jacob McCloskey}
\author[a]{Solange V. Ramirez}
\author[a]{William Schoenell}
\author[f]{Stephen A. Smee}
\author[a]{Jason E. Williams}

\affil[a]{Carnegie Science, The Observatories, 813 Santa Barbara Street, Pasadena, CA 91101, USA}
\affil[b]{Giant Magellan Telescope Organization, 300 N Lake Ave, 14th Floor, Pasadena, CA 91101}
\affil[c]{Turion Space, Irvine, CA, USA}
\affil[d]{LCS Optics LLC, Parkton, MD 21120, USA}
\affil[e]{The Ohio State University, Columbus, OH 43210, USA}
\affil[f]{Johns Hopkins University, Department of Physics and Astronomy, 3701 San Martin Drive,
Baltimore, MD 21218, USA}
\affil[g]{Capella Space, 438 Shotwell St., San Francisco, CA, 94110}
\affil[h]{Penn State University, University Park, PA 16802, USA}

\authorinfo{Further author information: (Send correspondence to M.C.)\\M.C E-mail: mcosens@carnegiescience.edu}

\begin{document} 
\maketitle

\begin{abstract}
The Magellan InfraRed Multi-Object Spectrograph (MIRMOS) is a planned next generation multi-object and integral field spectrograph for the 6.5m Magellan telescopes at Las Campanas Observatory in Chile. MIRMOS will perform $\rm R\sim3700$ spectroscopy over a simultaneous wavelength range of 0.886 - 2.404\um \, (Y, J, H, K bands) in addition to imaging over the range of 0.7 - 0.886\um. The integral field mode of operation for MIRMOS will be achieved via an image slicer style integral field unit (IFU) located on a linear stage to facilitate movement into the beam during use or storage while operating in multi-object mode. The IFU will provide an $\rm\sim18''\times26''$ field of view (FoV) made up of $\rm0.84''\times26''$ slices. This will be the largest FoV IFS operating at these wavelengths from either the ground or space, making MIRMOS an ideal instrument for a wide range of science cases including studying the high redshift circumgalactic medium and emission line tracers from ionized and molecular gas in nearby galaxies. We present here an update on the IFU design from our 2024 proceeding. Previously we utilized a re-imaging style slicer which required freeform pupil mirrors in order to achieve the required image quality while obeying significant packaging constraints near the instrument focal surface. In order to reduce the manufacturing cost and decouple the IFU from the configurable slit unit used in multi-object mode, we have moved the IFU deeper into the instrument, allowing for a switch to a virtual style IFU. This now requires a re-imaging doublet before the slicer mirrors, but removes the need for any freeform surfaces. We present here the optical design and predicted performance of the new MIRMOS IFU along with a conceptual design for the opto-mechanical system which will move the IFU between its active and stored positions.
\end{abstract}


\section{INTRODUCTION} \label{sec:intro} 

Wide-field optical integral field spectrographs (IFS) have been transformative for astronomical research and have thus seen significant investment at major observatories around the world (e.g., Keck/KCWI\cite{Morrissey2018}, VLT/MUSE \cite{Bacon2010}, Magellan/LLAMAS\cite{Furesz2020} or IFUM\cite{Mateo2022}, HET/VIRUS\cite{Hill2018}). These highly productive instruments have been particularly impactful in opening new windows to the resolved gas conditions in and around galaxies; allowing astronomers to study the gas morphology and kinematics in the circumgalactic medium (CGM), chemical (in-)homogeneity in the interstellar medium (ISM), and the strength of stellar feedback in differing environments. However, these investigations are often limited by the restriction to optical wavelengths with no equivalent wide-field IFS capabilities in the near-infrared (near-IR). Such instruments instead have followed adaptive optics systems and therefore prioritized spatial sampling at or near the diffraction limit rather than field coverage (e.g., VLT/ERIS\cite{Davies2023}, Keck/OSIRIS\cite{Larkin2006}, Gemini/NIFS\cite{McGregor2003}). This means that there are key questions in particular regarding the structure and gas content of the CGM and ISM that cannot be answered with our current capabilities.

\begin{figure}
    \centering
    \includegraphics[width=0.9\textwidth]{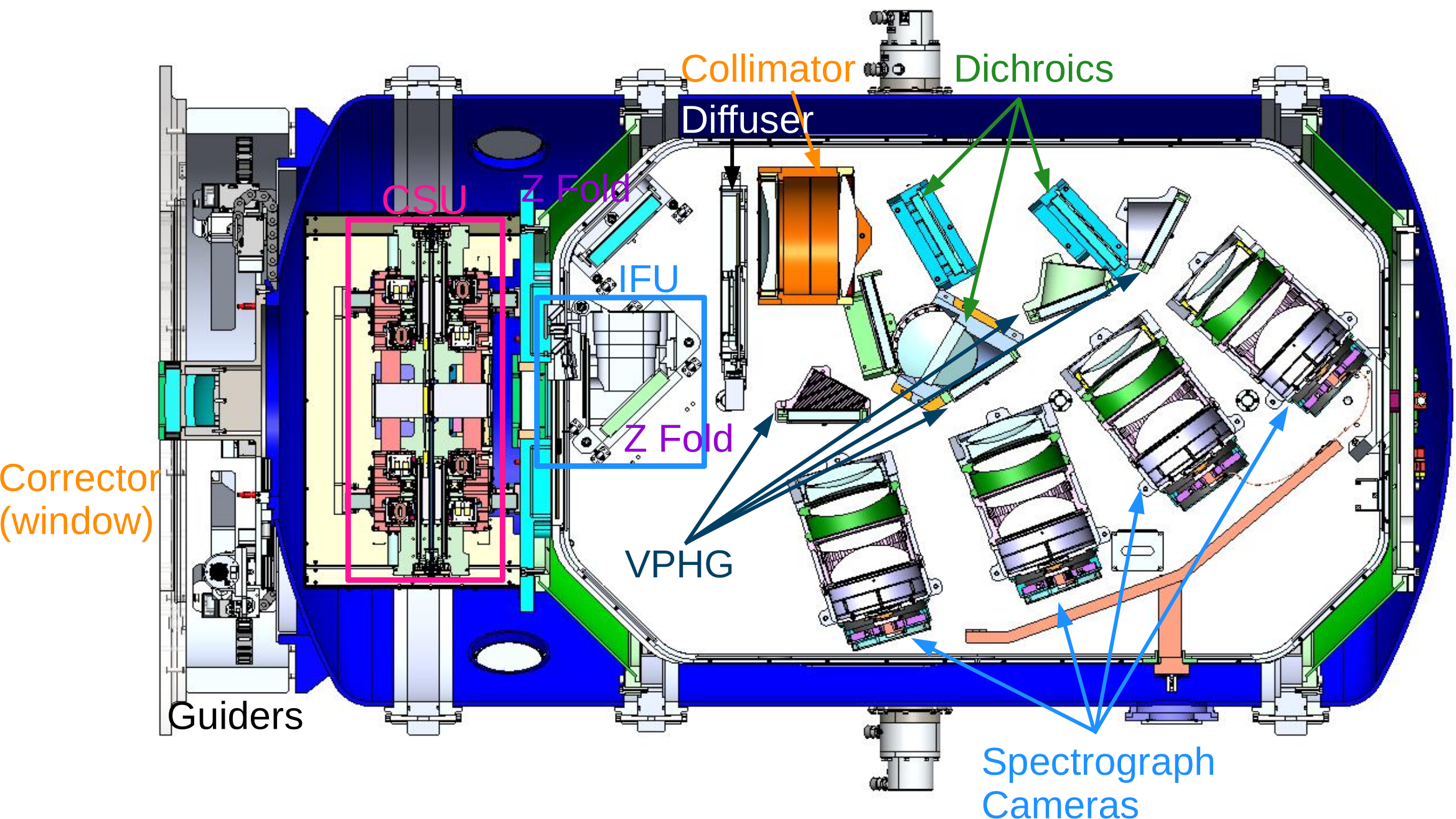}
    \caption{CAD model of the full MIRMOS instrument with key components labeled. Light enters the cryostat from the left through a two-element corrector with the front element doubling as the dewar window. Next, light passes through a slit mask formed by the CSU for long-slit or MOS mode, or the bars open for IFS mode. Then the light passes through a field lens between the CSU and IFU assemblies. If the IFU is not in use (as shown here), the optics of the IFU are moved out of the way and light is directed by the two Z Fold mirrors into the rest of the instrument. If the IFU is in use, these optics are moved into the beam as shown in Figure \ref{fig:IFUstage} and follows the lightpath in Figure \ref{fig:IFU_layout}. The rest of the optics are common between these two modes with the newly formed slits or pseudo-slits directed by the two Z Fold mirrors into the collimator, before being divided by a dichroic tree to the imaging camera (underneath the optical bench) or the four Y, J, H, and K gratings and cameras shown here. Each of these bands has a dedicated detector assembly allowing them to operate simultaneously.}
    \label{fig:MIRMOS}
\end{figure}

To address this critical capability, we are designing a slicer integral field unit (IFU) for the Magellan InfraRed Multi-Object Spectrograph (MIRMOS). The MIRMOS design is described in detail in\cite{Konidaris2020} with updates in\cite{Konidaris2022, Konidaris2024} and this conference\cite{Cosens2026} and summarized here. MIRMOS will be a next-generation multi-object and integral field spectrograph (MOS \& IFS) for the 6.5m Magellan telescopes with simultaneous $\rm R\sim3,700$ spectral coverage over 0.886-2.404\um \,(Y, J, H, and K bands) along with imaging from 0.7 - 0.886\um \, primarily for verifying target alignment. The MOS mode utilizes a configurable slit unit with 92 bar pairs to select targets over a $\rm13'\times3'$ field of view. For the IFS mode the bars will open and a periscope mirror will move into the beam to send light to the slicer IFU. Both modes utilize a common  collimator and dichroic tree to feed the five operating channels (four spectroscopic and one imaging). Each spectrograph channel will have a dedicated volume phase holographic (VPH) grating, camera, and Teledyne Hawaii-2RG detector. A CAD model of the instrument is shown in Figure \ref{fig:MIRMOS}.

A previous iteration of the IFU design was described in a 2024 SPIE proceedings\cite{Cosens2024}, but to reduce the cost and complexity of the optics we have performed a significant redesign. The previous system utilized freeform optics to produce the required image quality while respecting strict mechanical constraints imposed by other instrument components. In order to remove the need for freeform surfaces we have moved the system to a new location in the instrument and gone from a re-imaging style design to a virtual style IFU with a re-imaging doublet located before the slicer. In this proceedings we will describe this updated design in detail. The key science cases that produce the requirements for the MIRMOS IFS mode are discussed in Section \ref{sec:science}. The optical design of the new IFU system is described in Section \ref{sec:optics} with a focus on the changes since the 2024 proceedings and an effort to streamline the design process by scripting the ZOS-API. An update of the conceptual mechanical design to accommodate the changes to the system is shown in Section \ref{sec:mech}.

\section{SCIENCE DRIVERS} \label{sec:science}

MIRMOS is being designed as a facility instrument suitable for a wide range of science cases. The requirements for the IFU, however, are largely driven by the desire to (i) measure the morphology and kinematics throughout the CGM at cosmic noon, and (ii) probe the multi-phase ISM and stellar feedback in nearby star forming regions.

\subsection{Mapping Gas in the Circumgalactic Medium}

Understanding of the growth of galaxies over cosmic time requires understanding the surrounding gas in the CGM from which gas accretes onto galaxies to fuel new generations of star formation, and energetic outflows of enriched material also pass through and may terminate within the CGM. While quasar absorption-line spectroscopy has long provided a one-dimensional probe of this diffuse gas and its kinematics\cite{Chen2026}, the \textit{resolved} 3D structure of the CGM has only recently become observable with wide-field optical IFSs. These ground breaking observations have probed the Ly$\alpha$ nebulae around star-forming galaxies and QSOs at $2<z<3$, along with intra-group gas and extended inflows and outflows at $z < 1$ \cite{Johnson18, Chen19, Arrigoni2019,Leclercq2017,rup19, Vayner2023}.

However, because wide-field IFSs are currently restricted to the optical, CGM emission studies at $z > 1$ are largely limited either to weaker nebular transitions, reducing the sensitivity of the observations, or to resonant rest-UV transitions which makes interpretation of the data highly uncertain\cite{ver08,gro16}. Ly$\alpha$ and other UV resonance photons diffuse in space and in frequency, meaning that the observed spatial location and redshift distribution is not tied to the emission site\cite{lau09} making the emission mechanism, mass, and kinematics of the CGM gas poorly constrained by optical IFS observations. 

A wide-field near-IR IFS mode in MIRMOS would allow studies of resolved emission from the CGM of galaxies and QSOs at $z>1$ in strong nebular emission lines. These lines are optically thin, so their observed kinematics and emission morphologies directly translate to the underlying gas density and velocity fields. Combining such unambiguous morpho-kinematic maps with access to emission-line ratios dependent on gas density, ionization state, and metallicity, key advances can be made in our understanding of the gas cycles that govern galaxy evolution.

\subsection{Feedback and the Multiphase ISM in Nearby Galaxies}
The ISM is likewise a complex, multiphase environment. Diffuse atomic gas assembles into cold molecular clouds, which collapse to form stars \cite{peroux20}. These stars inject energy and momentum (``feedback'') into their surroundings in the form of radiation, winds, and supernovae; produce heavy elements; and recycle material back into the surrounding ISM \cite{mckee07, somerville15, thompson24}.  Which feedback sources are most effective at disrupting star-forming \hii \, regions and molecular clouds is a key open question in understanding the self-regulation and evolution of galaxies. Work with optical IFSs has begun to determine the energy budget for these feedback mechanisms in local galaxies, but crucial information in inaccessible with only optical spectra.

One reason is that the earliest stages of star formation occur within dense and heavily obscured molecular clouds, where young stars are deeply embedded in gas and dust \cite{mckee07}. These clusters provide direct insight into the initial phases of star formation and the early interaction between stars and their natal environments \cite{gregg24, pedrini24, rodrigez25}. Understanding the energy budgets in these clusters is critical, as initial studies seem to imply different feedback mechanisms dominate at such early times \cite{Olivier21}. Even at later evolutionary stages, studies with optical IFS do not agree with simulations on the importance of radiative versus mechanical feedback, particularly at low metallicity \cite{McLeod2019, Cosens2022, Jecmen2023}. To understand how stellar feedback operates and to obtain a complete census of star formation activity in galaxies, wide-field near-IR IFS observations are critical.

Further, the influence of these young stars cannot be understood without also tracing the influence of feedback on the molecular gas that fuels further star formation. MIRMOS will probe the molecular gas phase simultaneously and at the same spatial sampling as the ionized gas. The ro-vibrational \hh \, transitions throughout the near-IR spectral range probe both the gas content and excitation mechanism\cite{riffel13, kristensen23, peeters24}. MIRMOS observations across nearby galaxies will trace the extent and excitation of molecular gas simultaneously with ionized gas emission, providing a robust link between the sources of feedback energy in the \hii \, region and its impact on the gas reservoirs which feed further star formation.

\subsection{Requirements} \label{sec:req}

These key science cases drive the requirements for the MIRMOS IFU listed in Table \ref{tab:IFU_reqs} along with their design impact. Volume constraints and operational requirements are also listed here.

\begin{table}[h]
\centering
\caption{Key requirements for the MIRMOS integral field mode} 
\label{tab:IFU_reqs} 
\includegraphics[width=0.7\linewidth]{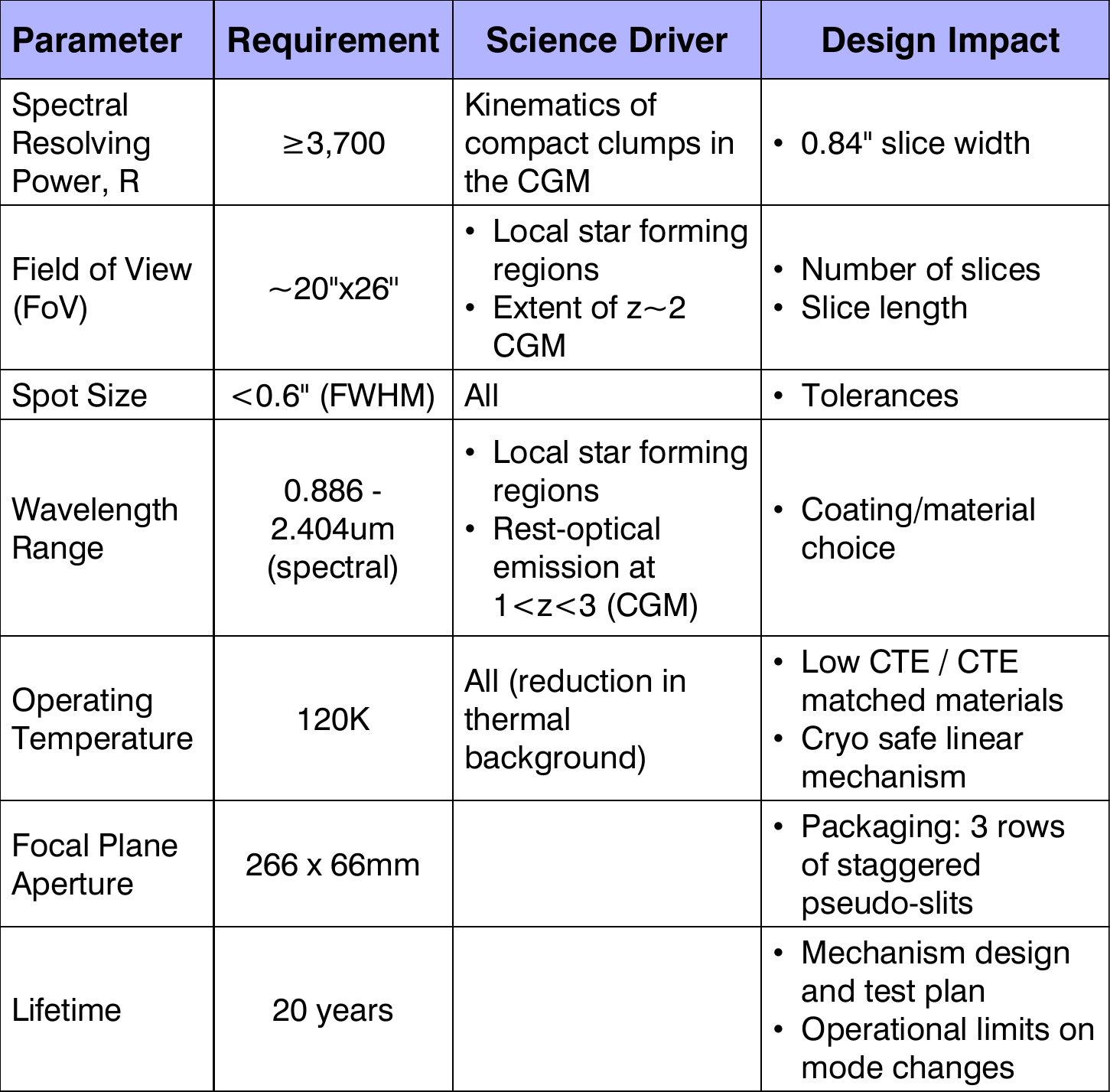}
\end{table}

Both of the key IFS science cases benefit from the largest FoV possible. However, another critical goal is to maximize the spatial and spectral resolution in order to disentangle the motion of distinct gas clumps in both the CGM and star forming regions. The chosen $\rm 0.84''$ slice width is the smallest that can be fully sampled at the detector, balancing the resolution and FoV. Additionally, this matches the nominal slit width to be used in the MOS mode for consistent spectral resolution.

There is also a requirement to be able to switch between the MOS and IFS modes within 30 minutes. However, the operational plan is to allow the use of only one mode per night. This is due to the need to take daytime calibrations. If this process can be done faster than expected or in cases where there is time to perform calibrations in the morning before engineering work must be completed the mechanism is being designed not to substantially limit a change in modes during the night.

\section{OPTICAL DESIGN} \label{sec:optics}
The previous version of the IFU design is described in detail in our 2024 proceeding\cite{Cosens2024} so will only be summarized here to provide context for the changes made. The original design required that the IFU be located in the front section of the instrument between the front corrector and the CSU. Since it was located before the slit, this design utilized a re-imaging style slicer IFU which formed pseudo-slits at the same location as the instrument focal surface when operating in MOS mode. The space available at this location for the optics and required mechanism was extremely constrained and so freeform pupil mirrors were required in order to achieve the required image quality. This resulted in both high expected manufacturing costs and tight tolerances on the alignment of the pupil mirrors. In order to simplify alignment and reduce the overall cost we have undergone a significant redesign by allowing the IFU to move after the focal surface and the CSU. It is now located after the field lens on the main optical bench of MIRMOS. This has necessitated moving a shutter that was going to be in this location to the prior location of the IFU, in front of the CSU. This change is minor and has no impact on the shutter design while reducing the load on the front of the CSU.

\begin{figure}[h]
    \centering
    \gridline{\fig{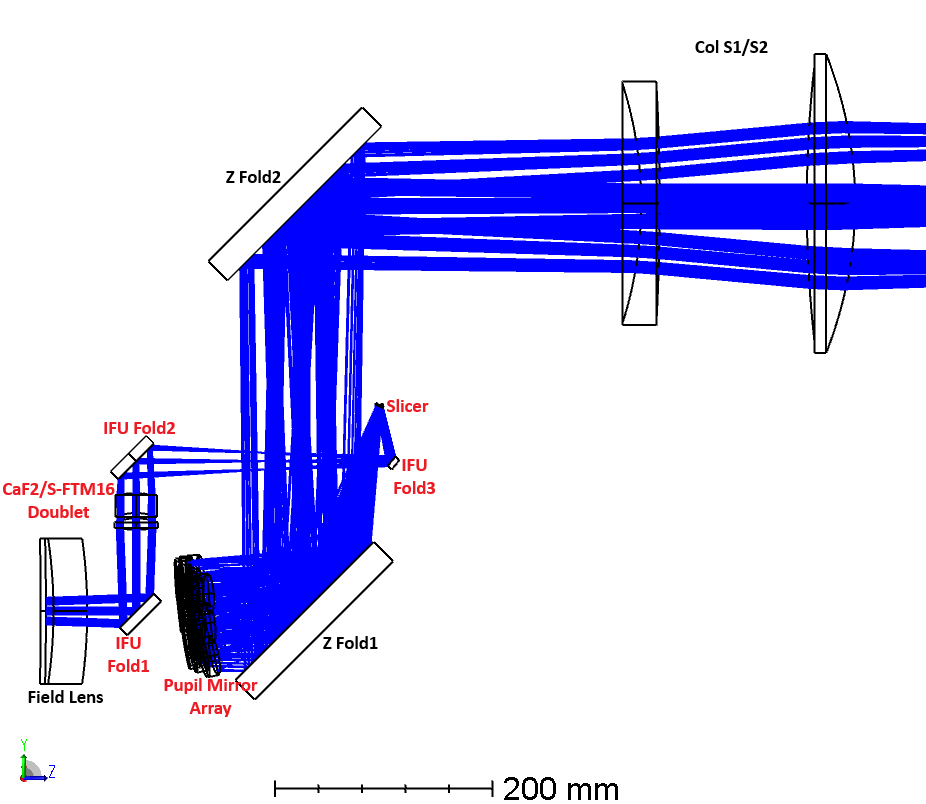}{0.48\textwidth}{}
            \fig{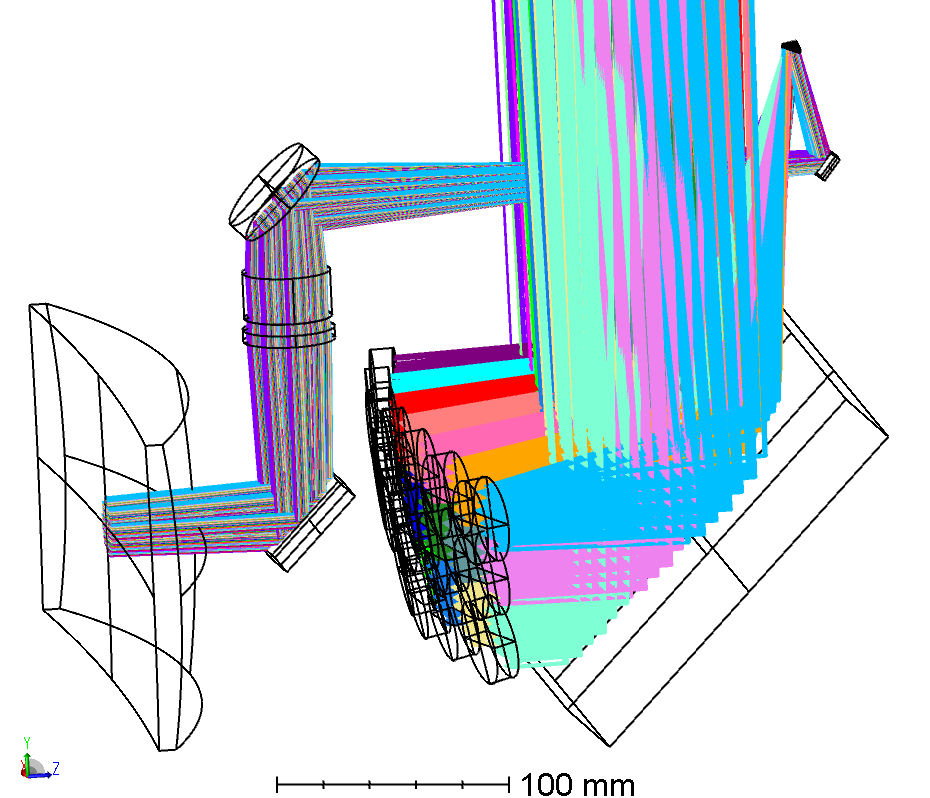}{0.48\textwidth}{}}
    \caption{Zemax model + raytrace of the IFU showing the location in the instrument (left) and the 21 unique slices and pupil mirrors (right) while in the engaged position. IFU Fold1 picks off the $\rm \sim~18''\times26''$ field. At this point the beam is diverging and must be focused at the slicer by the air-gapped CaF$_2$ / S-FTM16 doublet. Two additional fold mirrors direct the light to the slicer array which sends each spatial strip of the field to the 21 pupil mirrors. Light then goes to the common collimator and cameras.}
    \label{fig:IFU_layout}
\end{figure}

For this updated design, we utilize a virtual style IFU\cite{Weitzel96} located after the instrument focal surface and the field lens -- the first element of the collimator. This requires an air-gapped CaF$_2$/S-FTM16 doublet reimager to focus light on the slicer array. The slicer and pupil mirror arrays reformat the field into 21 discrete segments to be dispersed in wavelength by the spectrograph. Both arrays consist of curved (spherical) mirrors with a unique prescription and tip/tilt angles for each. Additionally, three flat fold mirrors are used to select the field of view of the IFU and package the assembly. The optical layout of the IFU is shown in Figure \ref{fig:IFU_layout}. 

In a traditional virtual style IFU the pupil mirrors would be flat. However, given the packaging constraints in MIRMOS and the location after the first element of the common collimator, slight curvature was required on the pupil mirrors. Still, since these surfaces are no longer freeform, the alignment tolerances become looser and manufacturing options become more flexible.

This system was designed iteratively in Zemax in order to achieve the desired performance while respecting the tight volume constraints. First, the center slice was designed with a parametric re-imager, collimator, and camera to get a first order approximation of the system. Then, the Field Lens and Zfold mirrors were added to provide references for the necessary mechanical constraints. The best position of the slicer from this iteration was used as a constraint for a standalone optimization of the re-imaging doublet. Next, the center slice of the IFU was optimized in the full system with the fixed re-imager. The rest of the IFU was built one slice at a time with constraints on the spacing between adjacent pupil mirrors and the gap between spectra on the detector. A final pass optimization was performed on the full IFU simultaneously to provide minor adjustments on the pupil mirror and spectra positions. This incremental optimization strategy was necessary as the optimizer in Zemax had difficulty converging with many free parameters unless the initial value was close to optimal.

The re-imaging doublet was optimized over the full spectral bandpass, but otherwise this optimization was performed in the K-band only as this provides the most stringent constraints on the pupil due to the need for masking thermal background. An evaluation of the system performance showed that the spot size requirements were being met at all but the outermost slices in this band. In order to evaluate the performance across all five of MIRMOS's bands, 105 unique configurations of slice position and band are required. To do this, automation of the analysis was required.

\subsection{Scripting the ZOS-API}

As discussed in Section \ref{sec:optics}, the IFU was optimized in only one band for simplicity. K-band was chosen since the Lyot stop is the most restrictive in this band. Since the IFU optics are reflective\footnote{Note: the reimaging doublet preceding the slicer is refractive, but this was separately optimized in the 4 spectral bands of MIRMOS (notably not in the imaging band)} performance in the K band should be representative of the others, but for verification of requirements and tasks such as making mock spectra to test reduction pipelines it is useful to verify performance across the full system. 
 
The ZOS-API appears to be a natural way to automate the generation of a full system model and subsequent analysis of the resulting performance by scripting these actions with Python. However, there is a large barrier to entry to utilizing this functionality and difficult to parse documentation. For this reason, IFU's are often designed and evaluated separately from the full systems they are integrated with. 

To evaluate the performance of the MIRMOS IFU, we have developed a series of Python scripts (available on GitHub\footnote{\url{https://github.com/mcosens/Zemax_IFU_tools}}) to incorporate the K-band IFU into the full MIRMOS model, evaluate the full system performance, and tolerance all slices. Since this model will change with the re-balancing of the collimator and cameras following cryogenic index of refraction measurements of the materials in use, the only practical way to do this was to make it automated and therefore straightforward to propagate later updates. 

\subsubsection{Performance Evaluation}
\begin{figure}[h]
    \centering
    \gridline{\fig{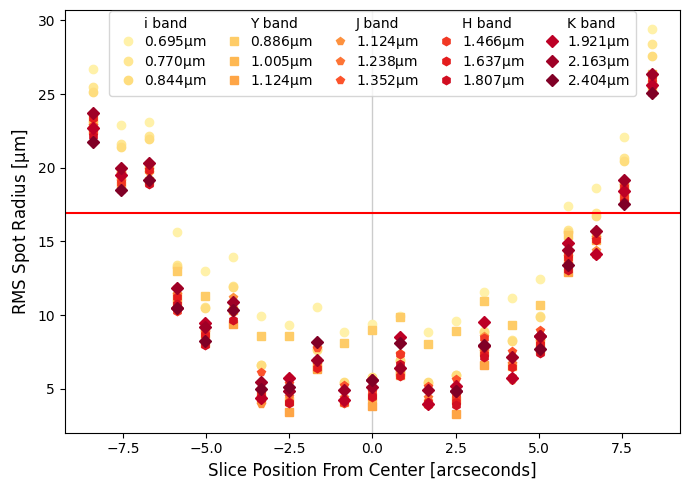}{0.45\textwidth}{(a)}
            \fig{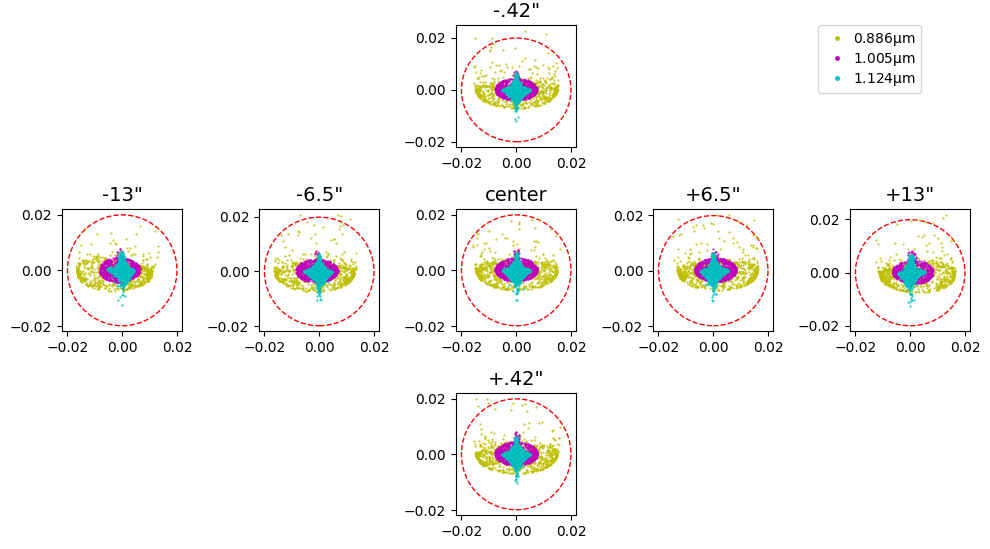}{0.55\textwidth}{(b)}}
    \caption{(a): RMS Spot radius as a function of slice position from the center of the IFU. Radii are averaged across the slice. Spots are colored by wavelength. The red line shows the $\rm FWHM<0.6''$ spot size requirement which is equivalent to $\rm16.9$\um \, RMS radius. This is met for all but the outer edges of the FoV. (b): Example spot diagram for the center slice in the Y band. The spot radii for these field positions are averaged to give a single value per slice shown on the left. The red circle highlights the same $\rm16.9$\um \, RMS radius requirement.}
    \label{fig:IFU_spots}
\end{figure}

\begin{figure}[h]
    \centering
    \includegraphics[width=0.9\linewidth]{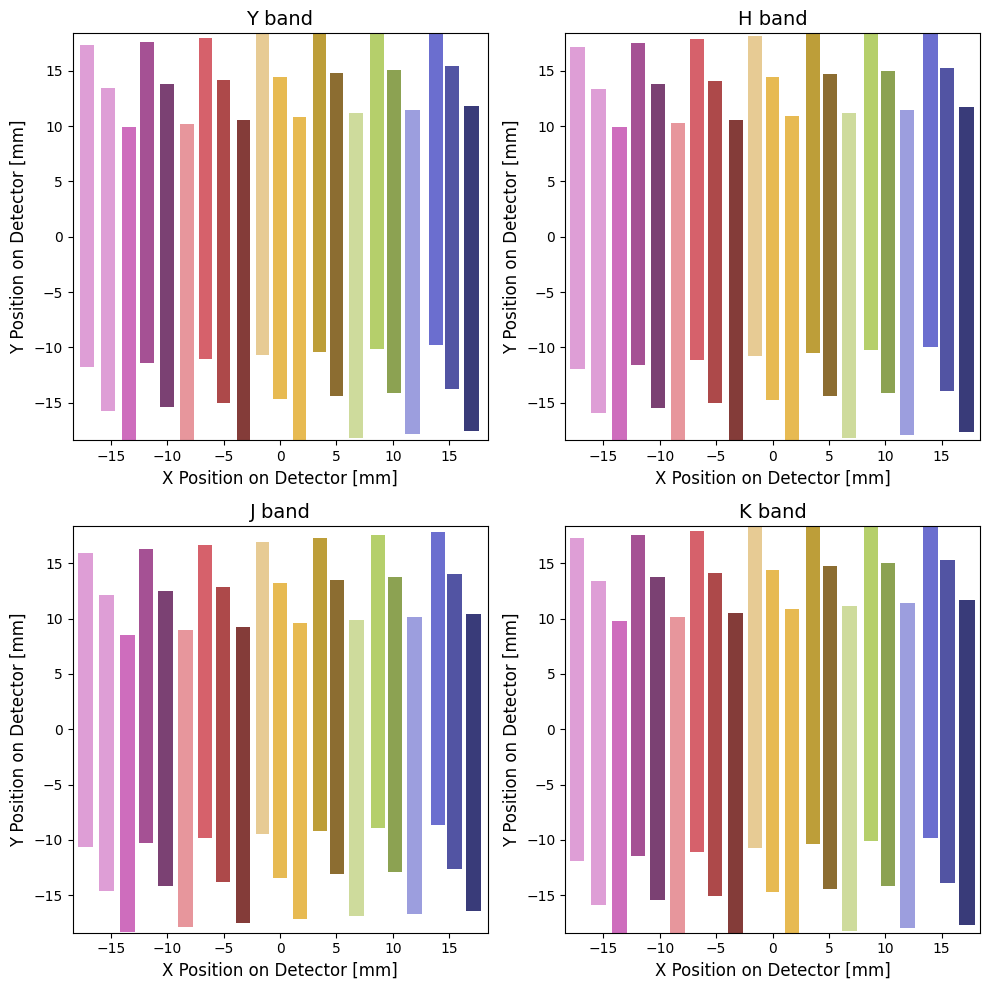}
    \caption{Footprint of each slice on the four spectrograph detectors showing sufficient space between each spectrum to differentiate the slices and extract spectra with a minimum gap of 10 pixels. Note that due to the staggering of the pupil mirror array, some slices experience a slightly truncated wavelength range in some bands ($\rm\lesssim4\%$ at maximum, see Figure \ref{fig:wave_vign}).}
    \label{fig:detector_footprint}
\end{figure}

\begin{figure}[h]
    \centering
    \gridline{\fig{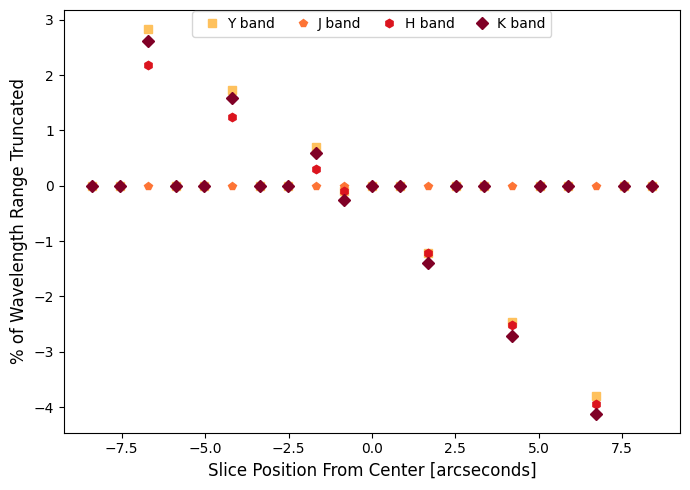}{0.48\textwidth}{(a)}
            \fig{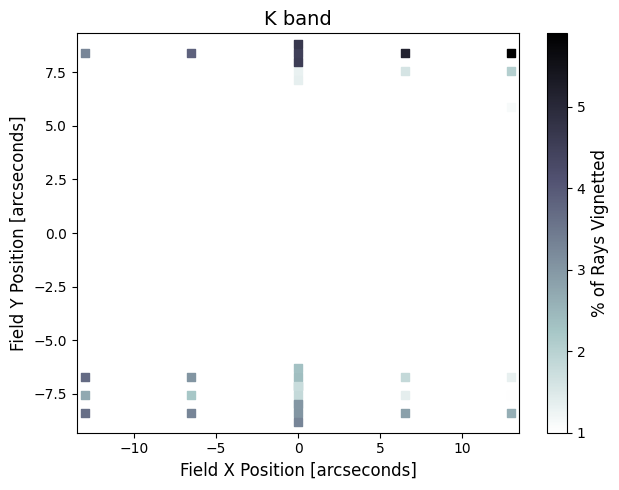}{0.48\textwidth}{(b)}}
    \caption{(a): Percentage of wavelength range truncated in each band as a function of position from the center of the IFU. Negative (positive) values indicate the red (blue) end of the spectrum falls off the detector. (b): 2D plot of vignetting in the K-band. All bands were evaluated but only K is shown here as is shows the largest vignetting fraction due to the more restrictive Lyot stop. This was evaluated only at the central wavelength for each band to avoid the impact of the spectra falling off the edge of the detector.}
    \label{fig:wave_vign}
\end{figure}

First, a script was written to combine the full MIRMOS optical model and the K-band only IFU. This produces a lens file with 105 unique configurations to be used in the subsequent analysis. Next, this full model is used for evaluating key performance metrics such as vignetting fractions, RMS spot radii, and spacing between spectra across the full wavelength range by looping through each configuration within Python and using the ZOS-API to perform the analysis normally done in the Zemax GUI. This not only allows the process to be automated but also for custom figures and simplified combination of results across configurations. We evaluate the RMS spot radius at 3 wavelengths per band and 7 field positions per slice - including the generation of spot diagrams for each (Figure \ref{fig:IFU_spots}), the extent and spacing of the spectra on each detector (Figure \ref{fig:detector_footprint}), the wavelength coverage of each slice (Figure \ref{fig:wave_vign}a), the vignetting fraction as a function of field position for each band (Figure \ref{fig:wave_vign}b), and the size and position of the pupil.

From this analysis we can see that the spacing between spectra at the detector is sufficient to perform spectral extraction at all bands, including the impacts of spot sizes and sloped spectra. The field dependent vignetting is $\rm\lesssim5\%$ everywhere, the RMS spot radius is consistent with the requirements across the vast majority of the field, and the pupil is well matched to the Lyot stop. There is some loss of wavelength coverage in the Y, H, and K bands for particular slices, but this is unavoidable due to the need for 3 rows of pupil mirrors. In all cases this is $\rm\lesssim4\%$ of the band.
 
\subsubsection{Tolerances}\label{sec:tol}
In order to evaluate the tolerances over the full IFU, we again script the ZOS-API to run a tolerance analysis on each slice individually and compile the results within Python to derive the overall tolerances. We include tolerances on the radius of curvature, centration, tip/tilt, thickness, and surface figure error for each of the slicer, pupil mirrors, fold mirrors, and re-imaging doublet. We allow compensators on the positioning of the re-imaging doublet (tip/tilt/decenter and space between fold mirrors). We run an inverse increment analysis, allowing each tolerance parameter to increase the RMS spot radius by 0.5\um \, (1\um \, was initially used as the increment, but this resulted in end-to-end images that did not meet the requirement in Table \ref{tab:IFU_reqs}). This method adjusts the tolerance parameters in the case that the initial values would exceed the limit applied. The adjusted parameters are compared across all slices and the same minimum adjustment allowed is used for all slices in subsequent steps.

\begin{SCfigure}[0.8][h]
    \includegraphics[width=0.5\textwidth]{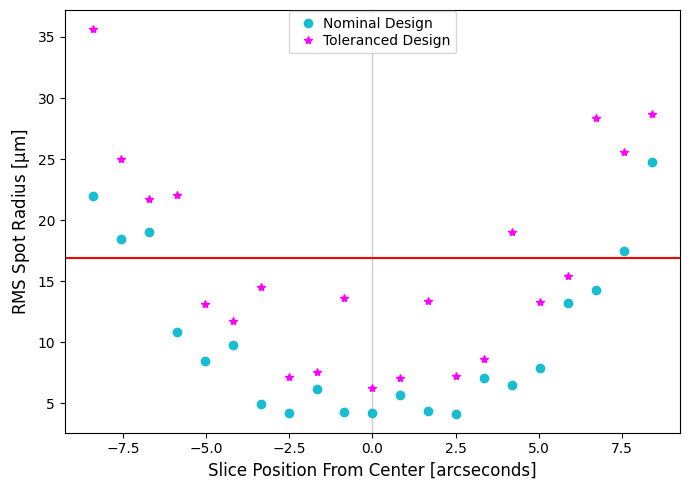}
    \caption{Results of tolerancing all optics in the IFU subsystem. This includes optical and opto-mechanical tolerances and a compensator on the position of the re-imaging doublet. The system is most sensitive to positioning tolerances, informing the need for adjustment. The resulting 80th percentile spot radii from a full Monte Carlo run of 500 iterations for each slice are shown in magenta with the nominal spots for reference (cyan). As in Figure \ref{fig:IFU_spots}, the requirement is shown by the red line. Currently, these tolerances have only been evaluated only at the center of K-band ($\rm2.163$\um). Future work will tolerance all spectral bands and explore areas to desensitize the design. \\}
    \label{fig:spots_tol}
\end{SCfigure}

Next, these new tolerance parameters are used to run a more in depth Monte Carlo analysis of their impact on the RMS spot radius. For each slice we run 500 iterations and evaluate the resulting spot size in the 50th and 80th percentile. The resulting spot sizes for the more conservative 80th percentile are shown in Figure \ref{fig:spots_tol} along with the nominal values from the ideal design. Currently, these tolerances have been evaluated only at the center of K-band ($\rm2.163$\um). Future work will tolerance all spectral bands and explore areas to desensitize the design to allow the system to meet the requirements with increased margin on alignment.

\section{MECHANICAL CONCEPT} \label{sec:mech}
\begin{figure}
    \centering
    \includegraphics[width=0.9\linewidth]{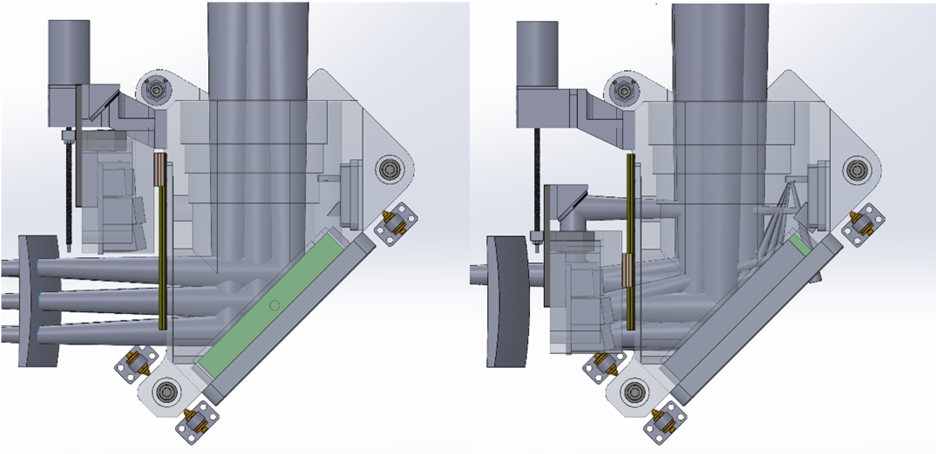}
    \caption{Conceptual design of the IFU mounting stage. The assembly is mounted to the large mirror ``Z Fold 1" that is common between the two modes. The moving portion of the stage holds ``IFU Fold1/2", the re-imaging doublet, and the pupil mirrors. This mechanism uses a common design and components (motor, rails, switches, detents) to the other MIRMOS mechanisms. The slicer and ``IFU Fold 3" are outside of the MOS beam shown on the left and therefore can remain fixed. The pushers shown will be used to adjust the tip/tilt of this entire assembly for instrument alignment and baffling is included in the semi-transparent portion of the assembly.}
    \label{fig:IFUstage}
\end{figure}

The IFU will consist of two separate assemblies, a mobile part holding the first two fold mirrors, re-imaging doublet, and pupil mirrors, and a fixed part holding the third fold, slicer array, and baffling. Both will be attached to the ``Z Fold 1" mirror base. The mobile part needs to move approximately 120\,mm to clear the optical path in MOS mode (IFU-off). The assembly and its two positions are illustrated in Figure \ref{fig:IFUstage}. Spring loaded detents that are engaged by default in the active position are base-lined in order to improve the repeatability of positioning. To simplify the design of this stage and increase the reliability of positioning we will impose an operational constraint that it only be moved at a single gravity vector. It is intended that the instrument not be changed between modes during the night due to the potential for a large amount of time needed for dedicated calibrations. At most, the mode may be changed once per night, meaning the time required for the switch will not have an appreciable impact on observing efficiency. If this is determined to be unnecessary later in the design and testing, this approach will be re-evaluated.

By grouping the IFU with the mount for ``Z Fold 1", it is possible to adjust the alignment of the whole system with respect to the rest of the MIRMOS instrument, enabling separate alignment of the IFU optics. From the preliminary tolerance, the slicer and third fold mirror positions can rely on dead reckoning while adjustment is required on the position of the re-imaging doublet. The complete pupil mirror array will be adjustable in tip and tilt (as a monolith, not individual pupil mirrors). This will align the IFU to Z Fold 1. Then to align the assembly with the rest of the system we will use the pushers shown in Figure \ref{fig:IFUstage} to align the pupil of both modes to each other and the pupil stop.

As discussed in our 2024 proceeding, this mechanism and its components will undergo rigorous testing including the ability to maintain accurate positioning at multiple orientations. Testing on the components of this stage have already begun as part of our prototyping efforts. These components are common to all of the linear mechanisms in MIRMOS and will soon undergo lifetime testing on a full prototype of the diffuser mechanism highlighted in Figure \ref{fig:MIRMOS}. The rails have already undergone an initial sanity check of functionality at cryogenic temperatures through testing in liquid nitrogen. Full testing will be performed in a new test chamber being delivered to Carnegie this summer with the ability to rotate automatically through a full 360$^\circ$ rotation. 

\section{SUMMARY} 
We have presented here an update the the optical design of the MIRMOS IFU first presented in our 2024 proceedings\cite{Cosens2024}. The original design utilized a re-imaging style slicer IFU with costly freeform optics to achieve the required image quality and has been replaced by a virtual style IFU with only spherical mirrors by allowing the system to move to a different location in the instrument and adding a re-imaging doublet. This design should prove more cost effective and easier to align within the instrument. 

In addition to these changes to the overall design we also developed a series of Python scripts which interface with the Zemax ZOS-API in order to systematically evaluate the system performance and tolerances over the full wavelength range and IFU field of view. These tools are publicly available and should serve as a reference for implementing similar functionality.

\acknowledgments 
We thank the entire MIRMOS team and our collaborators at CSEM, Teledyne, Universal Cryogenics, and INAF. This material is based on substantial funding from Carnegie Science. This research was funded by the Heising-Simons Foundation through grant 2021-2614. This material is based upon work supported by the National Science Foundation under Grant No. 2206374. We recognize generous support from the Ahmanson Foundation. The Mt. Cuba Astronomical Foundation funded MIRMOS. M.C. is supported by a Brinson Prize Postdoctoral Fellowship.

\bibliography{mirmos_ifu} 
\bibliographystyle{spiebib} 

\end{document}